\documentclass{mn}[3]
\usepackage{rotating}
\input psfig.sty
\begin{document}

\title[binary evolution \& blue stragglers]
{Primordial binary evolution and blue stragglers}
\author[Chen et al. ]{Xuefei Chen$^{1}$\thanks{
xuefeichen717@hotmail.com} and Zhanwen Han$^{1}$ \\
$^1$National Astronomical Observatories/Yunnan Observatory, CAS,
Kunming, 650011, P.R.China}
\maketitle

\begin{abstract}
Blue stragglers can significantly enhance the spectral energy
toward short wavelengths, especially in ultraviolet and blue
bands. Much evidence shows that blue stragglers are relevant to
primordial binaries. In this paper, we systematically studied blue
stragglers produced from primordial binary evolution via a binary
population synthesis approach, and examined their contribution to
the integrated spectral energy distributions of the host clusters.
The mass transfer efficiency, $\beta$, is an important parameter
for the final products (then blue stragglers) after mass transfer,
and it is set to be 0.5 except for case A binary evolution. The
study shows that primordial binary evolution may produce blue
stragglers at any given times and that different evolutionary
channels are corresponding for blue stragglers in different visual
magnitude regions (in V band) on the colour-magnitude diagram
(CMD) of clusters. The specific frequency of blue stragglers
obtained from primordial binary evolution decreases with time
first, and then increases again when the age is larger than 10Gyr,
while that from angular momentum loss induced by magnetic braking
in low-mass binaries increases with time and exceeds that of
primordial binary evolution in a population older than 3 Gyr.
Meanwhile, blue stragglers resulting from primordial binary
evolution are dominant contributors to the ISEDs in ultraviolet
and blue bands in a population between 0.3 and 2.0 Gyr. The value
of $\beta$ significantly affects on the final results, e.g the
specific frequency of blue stragglers decreasing with $\beta$,
blue stragglers produced from a high value of $\beta$ being more
massive, then contributing more to the ISEDs of the host clusters.
For old open clusters, the assumption of $\beta =1$ when the
primary is in HG at the onset of mass transfer matches the
observations better than that of $\beta =0.5$ from the locations
of BSs on the CMDs. Our study also shows that, for most Galactic
open clusters, the specific frequency of blue stragglers obtained
from our simulations is lower than that of observations, which is
puzzling. Nevertheless, primordial binary evolution cannot account
for {\it all} blue stragglers observed in old clusters.
\end{abstract}


\section{Introduction}
Blue stragglers (BSs) were first found by Sandage
\shortcite{sand53}. These stars have remained on the main sequence
for a time exceeding that expected from standard stellar evolution
theory. Many mechanisms, including single star models and binary
models, have been presented to account for the existence of BSs
(see the review of Stryker, 1993). At present, it is widely
believed that more than one mechanism plays a role for the produce
of BSs in one cluster and that binaries are important or dominant
for the production of BSs in open clusters and in the field
\cite{lan07,dal08,sol08}. Binaries may produce BSs by way of mass
transfer, coalescence of the two components, binary-binary
collision and binary-single star collision. The collision of
binary-binary or binary-single may lead binaries to be tighter or
farther apart. So a binary will advance or prolong mass transfer
after collision, which may affect BS birth rate in a cluster.
These collisions are relevant to dynamics and environment in the
host cluster. Some works taking account of effects of collisions
\cite{hur01,hur05,gle08} showed a significant importance of
collisions for BS production in M67. However, in low density open
clusters or in the field, the dynamical interaction should be
small and have little contribution to BSs. In this paper, we are
only concerned with BSs resulting from the evolutionary effect of
primordial binaries, i.e. mass transfer and coalescence of two
components.

Binary evolution has been well studied in the last 50 years, and
it has been successfully used to explain many observed weird
objects, such as Algols, Symbiotic stars, cataclysmic variables,
type Ia supernovae etc. Here we list some recent reviews on binary
evolution and its application: van den Heuvel 2006, Webbink 2006,
Bethe et al. 2007, Tutukov \& Fedorova 2007, Langer \& Petrovic
2008. In recent years, binary evolution has been applied in
stellar population synthesis to study the characteristics of
clusters and galaxies. For example, Han et al.\shortcite{han07}
provided a binary model for the UV-upturn of elliptical galaxies.
Binaries differ from single stars in evolution due to mass
transfer between the two components. The less massive component
may accrete material from the more massive (or more evolved) one,
and some strange phenomena appear as above in observations.
Moreover, the two components may merge to become a single star. FK
Com is believed to be a typical system which is in the final stage
of coalescence \cite{str93}. Binary mass transfer was originally
advanced by McCrea \shortcite{mc64} to explain the BS phenomenon.
A number of attempts have been made to test this hypothesis in
open clusters ever since, from both detailed binary evolution and
Monte Carlo simulations
\cite{pol94,hur01,hur05,chen04,chen08a,chen08b,and06,tian06}.
These studies show that primordial binary evolution (PBE),
including mass transfer and coalescence of two components, are
important for BS formation in some open clusters. However, these
works are not enough to studying clusters from stellar population
synthesis.

Recently, Xin and her collaborators \shortcite{xin05,xin07}
investigated BSs in Galactic open clusters and examined their
contributions to the integrated spectral energy distributions
(ISEDs) of the host clusters. Their studies showed significant
enhancements toward short wavelengths, especially in ultraviolet
(UV) and blue bands. Their work provided a general picture of BSs
in Galactic open clusters from observations. In this paper, we
will systematically study the production of BSs from PBE using
binary population synthesis and show their contribution to the
host clusters in theory. Using this result, we may study the role
of binary evolution of BS production in Galactic open clusters by
way of comparing our results with observations.

We describe the binary population synthesis (BPS) model in section
2 and show the simulation results in section 3. In section 4, we
selected some Galactic open clusters to compare with our
calculations. The summary is shown in section 5.

\section{Evolutionary Channels and the binary population synthesis model}

\subsection{Evolutionary Channels}
According to the evolutionary state of the primary at the onset of
mass transfer, three mass transfer cases are defined, i.e. case A
for the primary being on the main sequence, case B for the primary
after the main sequence but before central He burning, and case C
for the primary during or after central He-burning \cite{kip67}.
The secondary, as mentioned in section 1, may accrete material
during Roche lobe overflow (RLOF) or merge with the primary into a
single star after RLOF. The mass ratio, $q=M_1/M_2$, of the two
components at the onset of mass transfer is an important parameter
in binary evolution. Here $M_1$ and $M_2$ are the masses of the
mass donor and the accretor, respectively. If $q$ is lower than a
critical value, $q_{\rm crit}$, the mass transfer is stable. The
secondary goes upward along the main sequence in response to
accretion, if it is a main sequence star, and becomes a BS when it
is more massive than the turnoff of the host cluster ({\it Channel
I}). All of the three cases above may produce BSs in this way but
in various orbital period ranges \cite{chen08b}. It is also likely
that the secondary fills its Roche lobe during RLOF and the system
becomes a contact binary. Previous studies indicate that this
contact binary will eventually coalesce as a single star
\cite{web76,egg00,li05}, although there are some controversies
over the timescale of coalescence
\cite{eggen89,vant94,dry02,bil05}. If both components are on the
main sequence, their remnant is also a main-sequence star (i.e.
case A mass transfer) and evolves in a way similar to a normal
star with that mass. So the remnant may be a BS if it is more
massive than the turnoff ({\it Channel II}). If $q>q_{\rm crit}$,
mass transfer is dynamically unstable, and a common envelope(CE)
is formed. The CE may be ejected if the orbital energy deposited
in the envelope overcomes its binding energy, or the binary will
merge into a single star. If the two components are main sequence
stars, the remnant of coalescence will still be on the main
sequence, and it is a BS if its mass is beyond the turn-off mass
of the host cluster ({\it Channel III}). Binary coalescence of a
contact binary or dynamically unstable RLOF (from case A mass
transfer) is a popular hypothesis for single BSs
\cite{mat90,pol94,and06,chen08a}.

\subsection{The binary population synthesis code}
We employ the binary population synthesis code originally
developed by Han et al. in 1994, which has been updated regularly
ever since
\cite{han94,han95,han95b,han98,han02,han03,han04,han07}. With this
code, millions of stars (including binaries) can be evolved
simultaneously from the zero-main-sequence to the WD stage or a
supernova explosion. The main input of the code is a grid of
stellar evolution models, which are calculated from Eggleton's
stellar evolution code \cite{egg71,egg72,egg73,han94,pol95,pol98}.
In this paper, we use a Population I model grid ($Z=0.02$),
including the evolution of normal stars in the range of $
0.08-100.0M_\odot$ with hydrogen abundance $X=0.7$ and helium
abundance $Y=0.28$ and helium stars in the range of
$0.35-0.75M_\odot$. Single stars are evolved via interpolations in
the model grid.

The main process for binary evolution we are concerned with is the
first mass transfer as described in section 2.1. For this mass
transfer, we adopt $q_{\rm crit}=3.2$ when the primary is on the
main sequence or in the Hertzsprung gap. This value is supported
by detailed binary evolution calculations
\cite{han00,chen02,chen03}. If the mass donor is on the first
giant branch or AGB, $q_{\rm crit}$ is obtained by setting
$\zeta_{\rm S}=\zeta_{\rm L}$, where $\zeta_{\rm S}$ and
$\zeta_{\rm L}$ are the adiabatic mass-radius exponent and Roche
lobe mass-radius exponent of the mass donor (see  Hjellming \&
Webbink \shortcite{hje87} for details), respectively, and they are
fitted from the data of numerical calculations of Hjellming \&
Webbink \shortcite{hje87} and Soberman et al.
\shortcite{sober97}(see also Han et al., 2001)(here after we use
HW87 to present $q_{\rm crit}$ obtained from this way). Recently,
some fully binary calculations \cite{han02,chen08a} have
demonstrated another expression of $q_{\rm crit}$ for mass
transfer between a giant and a main sequence star. As an
alternative, we also adopt the results of Chen \& Han
\shortcite{chen08b}(CH08 hereinafter) to examine the consequence
of varying this value. The two descriptions of $q_{\rm crit}$
differ from each other. For instance, $q_{\rm crit}$ of HW87
results from a polytropic model and depends on the core mass
fraction of the mass donor at the onset of RLOF and mass transfer
efficiency during the RLOF, while that of CH08 is from detailed
binary evolution calculation and depends on stellar radius,
stellar mass and also on mass transfer efficiency. In CH08, the
more evolved a star, the less stable the RLOF, while it is the
opposite in HW87.

We assume that the mass transfer is always conservative for case A
evolution, while 50 per cent of the matter from the primary is
lost from the system for other cases. The mass lost from the
system takes away a specific angular momentum in units of the
specific angular momentum of the system \cite{ph92}. We have not
included angular momentum loss (AML) prior to RLOF. However, AML
of low-mass binaries induced by magnetic braking is important for
BS formation in old clusters. We examined its contribution to BSs
in another way (see section 2.3) when the age is older than 1 Gyr.

The problem of the CE ejection criterion is still open at present.
Here we introduce a classic $\alpha$-formalism, including two
model parameters, $\alpha_{\rm CE}$ for the CE ejection efficiency
and $\alpha_{\rm th}$ for the thermal contribution to the binding
energy of the envelope. The CE is ejected if
\begin{equation}
\alpha _ {\rm CE} \Delta E_{\rm orb} >E_{\rm gr}-\alpha _{\rm th}
E_{\rm th},
\end{equation}
where $\Delta E_{\rm orb}$ is the orbital energy released from
orbital decay, $E_{\rm gr}$ is the gravitational binding energy
and $E_{\rm th}$ is the thermal energy of the envelope. Both
$E_{\rm gr}$ and $E_{\rm th}$ are obtained from full stellar
structure calculations \cite{han94,dewi00} instead of analytical
approximations. We adopt $\alpha_{\rm CE}=\alpha_{\rm th}=0.75$ or
1.0 in our model calculations \footnote{As an alternative,
Nelemans, Verbunt and Yungelson \shortcite{nel00} and Nelemans \&
Tout \shortcite{nel05} suggested a $\gamma$-algorithm from the
view of the angular momentum balance to describe the CE evolution,
which may explain the formation of all kinds of close binaries.
However, as argued by Webbink \shortcite{web07}, {\it the
significance of this finding is itself open to debate }(cf.
Webbink, 2007).}.

In order to obtain the colours and the SED of the populations
produced in our simulations, we use the latest version of the
comprehensive BaSeL library of theoretical stellar spectra
\cite{lcb97,lcb98}, which gives the colours and SEDs of stars with
a wide range of metallicity $Z$, stellar surface gravity ${\rm
log}g$ and effective temperature $T_{\rm eff}$.

\subsection{Angular Momentum Loss in Low-mass Binaries }
The recent study of Chen \& Han \shortcite{chen08a} demonstrated
that AML is likely a main factor leading to BS formation in the
old open cluster M67, indicating that AML should not be ignored in
old clusters. We roughly estimated the AML contribution to BS
formation in a similar way to Chen \& Han \shortcite{chen08a}. A
semi-empirical formula for the orbital period variation is adopted
here \cite{ste06}:
\begin{equation}
$${\rm d} P_{\rm orb} \over {\rm d} t$$ = -(2.6+1.3) \times 10^{-10}P_{\rm orb}^{-1/3}e^{-0.2P_{\rm orb}}
\end{equation}
where $P_{\rm orb}$ is in days and time in years. For very short
orbital periods the exponential factor is close to unity and
varies very little during the subsequent evolution of the orbital
period of the binaries, and consequently, it is ignored.

We will firstly find out the time, $t_{\rm RLOF}$, at which the
primary fills its Roche lobe for low-mass binaries. The Roche lobe
of the primary $R_{\rm cr1}$ is calculated by \cite{egg83}
\begin{equation}
R_{\rm cr1}/A=$$ 0.49q^{2/3} \over 0.6q^{2/3}+{\rm
ln}(1+q^{1/3})$$ ,
\end{equation}
where $A$ is the separation and $q=M_{\rm 1}/M_{\rm 2}$.

The timescale from the onset of RLOF to being contact, and to the
final coalescence are ignored here, since AML may lead the systems
to reach contact, and then to coalesce very quickly (see Chen \&
Han, 2008a). Meanwhile, due to the very small evolution of the two
components prior to RLOF, the remnants of coalescence are ZAMS
stars with a total mass equal to the sum of the two components. If
$t_{\rm RLOF}$ is less than the cluster age $t_{\rm c}$, we then
evolve the mergers to $t_{\rm c}$ and obtain the characteristics
of the mergers at $t_{\rm c}$.

\subsection{Monte Carlo simulation parameters}
To investigate BSs from mass transfer and binary coalescence, we
performed a series of Monte Carlo simulations for a sample of
$10^5$ binaries (very wide binaries are actually single stars). A
single starburst is assumed in the simulations, i.e. all the stars
have the same age and metallicity ($Z=0.02$). The initial mass
function (IMF) of the primary, the initial mass ratio distribution
and the distribution of initial orbital separation are as follows:

i) the IMF of Miller \& Scalo \shortcite{ms79} is used and the
primary mass is generated from the formula of Eggleton, Fitchett
\& Tout \shortcite{egg89}:
\begin{equation}
M_{\rm 1}=$$0.19X\over (1-X)^{0.75}+0.032(1-X)^{1/4}$$
\end{equation}
where $X$ is a random number uniformly distributed between 0 and
1. The mass ranges from 0.8 to $100M_\odot$.

ii) the mass ratio distribution is quite controversial. We include
three different mass ratio distributions in the simulations. One
is a constant distribution \cite{maz92} as
\begin{equation}
n(q')=1,  0\le q' \le 1
\end{equation}
where $q'=1/q=M_2/M_1$. An alternative is a rising distribution
\begin{equation}
n(q')=2q',  0\le q' \le 1
\end{equation}
or the third case where both components are chosen randomly and
independently from the same IMF (uncorrelated).

iii) We assume that all stars are members of binary systems and
the distribution of separations is constant in log$a$ ($a$ is
separation).
\begin{equation}
an(a)=\left\{
\begin{array}{ll}
\alpha_{\rm sep}(a/a_0)^m, &a \le a_0\\
\alpha_{\rm sep}, & a_0<a<a_1\\
\end{array}
   \right.
\end{equation}
where $\alpha =0.070, a_0=10R_{\odot},a_1=5.75\times
10^6R_{\odot}=0.13{\rm pc}$ and $m=1.2$. This distribution gives
an equal number of wide binary systems per logarithmic interval
and 50 per cent of systems are with orbital periods less than 100
yr \cite{han95}.

\section{Simulation Results}
We performed five sets of simulations (see Table 1) for a
Population I composition ($X=0.70$, $Y=0.28$ and $Z=0.02$) to
systematically investigate the BS formation from PBE. The age
ranges from 0.1 to 20 Gyr. The first set is a standard set with
the best choice model parameters from the study of the formation
of hot subdwarfs (Han et al. 2002, 2003). We vary the model
parameters in the other sets to examine their influences on the
final results \footnote{Note that AML has not been included in any
of the five sets. The contributions of AML are calculated
separately from PBE by the method of section 2.3.}.

\begin{table}
 \begin{minipage}{80mm}
 \caption{Simulation sets (metallicity $Z=0.02$) in this paper.$n(q')$ = initial mass ratio distribution;
 $q_{\rm c}$ = the critical mass ratio with the first RLOF in FGB or
 AGB; $\alpha_{\rm CE}$ = CE ejection efficiency;
 $\alpha_{\rm th}$ = thermal contribution to CE ejection.}
 \label{tab1}
   \begin{tabular}{ccccc}
\hline
Set &$n(q')$&$q_{\rm c}$&$\alpha_{\rm CE}$&$\alpha_{\rm th}$\\
\hline
1 &constant&HW87&0.75&0.75\\
2 &Uncorrelated&HW87&0.75&0.75\\
3 &Rising&HW87&0.75&0.75\\
4 &constant&CH08&0.75&0.75\\
5 &constant&HW87&1.0&1.0\\
\hline \label{c-fgb}
\end{tabular}
\end{minipage}
\end{table}

\begin{figure*}
\centerline{\psfig{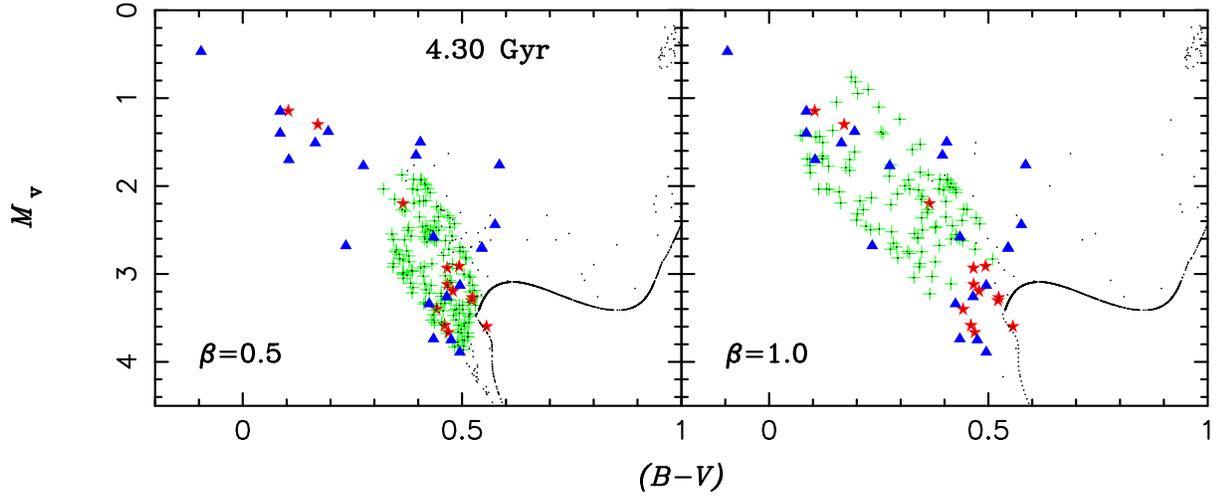}}
\caption{Colour-magnitude diagrams when the population is at
4.3Gyr. Here $\beta$ is the mass fraction of the matter lost from
the primary accreted by the secondary when the primary is in HG at
the onset of RLOF. The green crosses and red stars are for BSs
from mass transfer and binary coalescence, respectively, and the
small black dots are for other objects in the population. The
triangles are observed BSs of M67 from Sandquist et al. (2003).  }
\label{m67}
\end{figure*}

From our simulation, PBE may produce BSs at any given time. For
the standard set, we plotted colour-magnitude diagrams (CMDs) such
as Fig. \ref{m67} of the population at all ages we studied and
obtained some whole properties of BSs from PBE. Due to the size
limit of this paper, we have not shown other CMDs here and just
describe some whole properties as follows. One may send a request
to xuefeichen717@hotmail.com for these figures.

\begin{figure*}
\centerline{\psfig{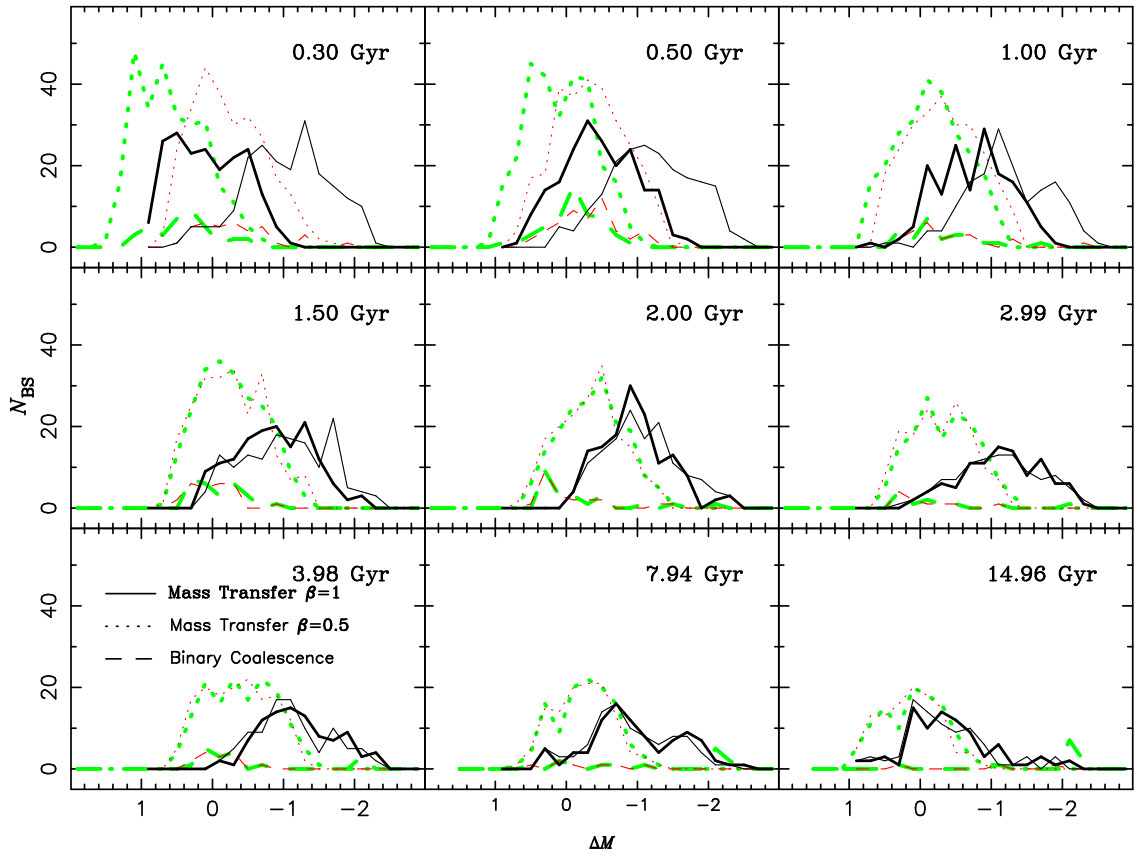}}
\caption{Distribution of BSs around the turnoff from primordial
binary evolution for different channels (i.e. mass transfer and
binary coalescence). The abscissa is magnitude difference between
BSs and the turnoff. The thick lines are for visual magnitude
(i.e. after bolometric corrections) and the thin ones are for
bolometric magnitude. Note that BSs from AML are not shown in this
figure.} \label{nd}
\end{figure*}

\subsection{Distribution on colour-magnitude diagrams}
Fig. \ref{nd} shows the distributions of BSs around the turnoff at
various ages from both mass transfer ({\it Channel I}, the dotted
lines) and binary coalescence ({\it Channels II and III}, the
solid lines). The abscissa is magnitude difference between BSs and
the turnoff, that is, $\Delta M=M_{\rm BS}-M_{\rm to}$, where
$M_{\rm BS}$ and $M_{\rm to}$ are the bolometric (or visual)
magnitudes of the BS and of the turnoff, respectively. Two sets of
magnitude difference (bolometric and visual) are presented in this
figure to show the difference between the intrinsic and observed.
Due to high bolometric corrections for massive stars, BSs move
toward high-magnitude regions in comparison to the turnoff after
bolometric correction when the population is younger than 1Gyr.

From Figs. 1 and 2 we see that, many BSs are below but bluer than
the turnoff. They may extend into the region 1 mag (hereinafter
mag or magnitude means visual magnitude without expression) lower
than the turnoff. These objects are mainly from {\it Channels I
and III}. They are less evolved than the turnoff and may
contribute more to the flux of V band in the following evolutions.
Most BSs from {\it Channels I and III} are within 1.5 mag of the
turnoff, while some ones from {\it Channel II} may have magnitudes
lower than the turnoff by about 2.3 mag. For {\it Channels I and
III}, the secondaries either accrete only a part of the primary's
mass or have very low initial masses (because of large $q$), while
the secondaries undergoing {\it Channel II} have relatively high
initial masses and the sum mass of two components is conserved for
the remnants. Thus, BSs from {\it Channel II} are generally more
massive than those from {\it Channels I and III}. However, {\it
Channel II} cannot produce BSs 3 mag brighter than the turnoff.
Extremely high luminous BSs have been observed in some open
clusters, e.g F81 in old open cluster M67. It is believed that F81
is likely from triple (or more) stars. Whatever, PBE without
dynamical interaction cannot account for {\it all} BSs even in
open clusters.

Between 0.4 and 1.5 Gyr, BSs with high magnitudes (i.e very close
to the turnoff) scatter all over the main sequence and the spread
decreases with increasing mass. This phenomenon gradually
disappears with time, and eventually, the spread of BSs on the
main sequence is similar for various masses. The time scales of
BSs in different magnitude regions at various ages can account for
this. For BSs near the turnoff, they have relatively longer
main-sequence time scales than those further away from the
turnoff, especially when the population is young. Thus, some BSs
near the turnoff are likely very close to zero-age main sequence
and some BSs formed earlier move to the red side of the main
sequence, resulting in a large spread around the turnoff. However,
it is different for massive BSs. Since the population is young,
massive BSs have very short main-sequence timescales and so is
unlikely to stay close to the zero-age main sequence, leaving gaps
in these regions. With increasing age, BSs with relatively high
mass are also likely to be close to the ZAMS, and then the scatter
is similar for BSs with various masses.

The effect of contact binaries ({\it Channel II}) on BSs is
intermittence. For example, BSs formed in this way appear at one
time and disappear for the following time (i.e. the earlier BSs
evolve off the main sequence while no new BSs are produced), and
then appear again later. The intermittence comes from the very
short time scales for the BSs formed in this way. Due to the
relatively high mass, BSs from this way (if they exist) are main
contributors to the ISED of the host clusters. As we compare our
results with M67, some observed BSs are obviously located in the
region from this channel (see Fig. \ref{m67}), indicating {\it
Channel II} having an effect on BS formation in this cluster,
although the effect might be small.

\subsection {The Specific Frequency}
\begin{figure}
\centerline{\psfig{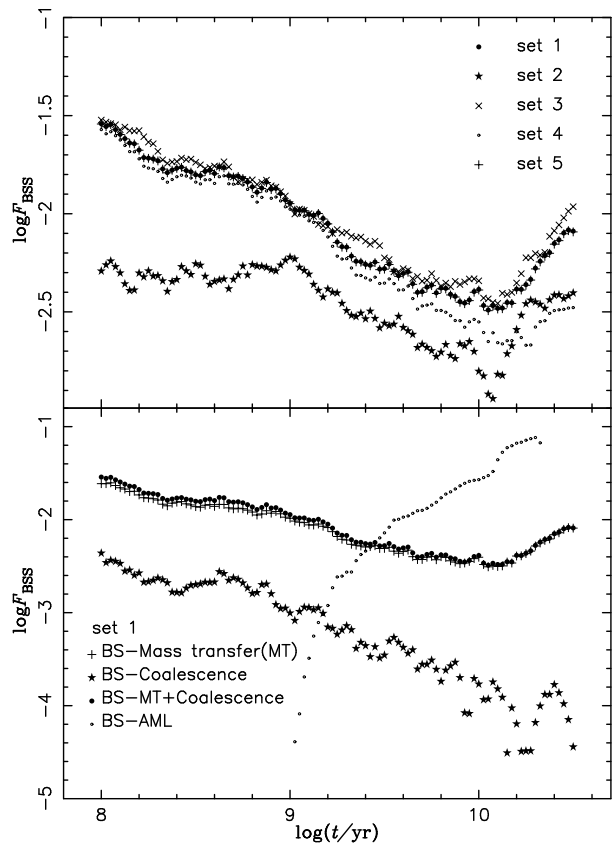}}
\caption{Specific frequency of BSs from primordial binary
evolution for different simulation sets (the upper panel, without
AML) and for different channels in the standard simulation set
(set 1, the bottom panel). The result of set 2 is obviously
different from the others since we adopted an {\it uncorrelated}
initial mass ratio distribution.} \label{FBSS}
\end{figure}

The BS number from the simulations, $N_{\rm BS}^{\rm b}$, depends
slightly on the population age in our simulations, while the
specific frequency, ${\rm log}F_{\rm BSS}$($ \equiv {\rm
log}(N_{\rm BS}^{\rm b}/N_2^{\rm b})$, $N_{\rm 2}^{\rm b}$ is the
number of stars within 2 mag below the main-sequence turnoff),
heavily depends on the age due to the increase of $N_{\rm 2}^{\rm
b}$, as shown in Fig. \ref{FBSS}, which includes the results from
different calculation sets (the upper panel, no AML having been
included) and from different evolutionary channels in the standard
simulation set (set 1, the bottom panel).

From Fig. \ref{FBSS} we see that, in all of the five sets, ${\rm
log}F_{\rm BSS}$ decreases with time first, and then increases
when the age, $t$, is larger than 10 Gyr. The decrease of ${\rm
log}F_{\rm BSS}$ before 1.5 Gyr comes from the increase of $N_{\rm
2}^{\rm b}$. After that, the number of potential binaries which
may contribute to BSs decreases, leading to fewer BSs formed. On
the other hand, $N_2^{\rm b}$ continues increasing. Thus, ${\rm
log}F_{\rm BSS}$ continues decreasing. With increasing time, the
primaries in long-orbital-period binaries gradually enter into AGB
phase, dramatically expand, and some of them may fill their Roche
lobe and start RLOF. Due to large stellar winds in the AGB phase,
these mass donors at the onset of RLOF are probably much less
massive than before and this mass transfer is easily stabilized,
resulting in some long-orbital-period BSs. In the standard set,
the long-orbital-period BSs first appear at 2.24Gyr, and increase
in number with time, since more primordial binaries with
long-orbital periods become contributors to BS numbers with time.
As a consequence, ${\rm log}F_{\rm BSS}$ begins to increase when
${\rm log} (t/{\rm yr})>10$.

As shown in Fig. \ref{FBSS}, the distribution of the initial mass
ratio is the main factor affecting the BS specific frequency,
since both the onset time and the stability of RLOF are relevant
to initial mass ratios. The value of ${\rm log}F_{\rm BSS}$
depends slightly on the critical mass ratio for dynamically stable
RLOF, especially when ${\rm log}(t/{\rm yr})>9.2$, when binaries
with long-orbital periods start contributing to BS formation. Both
$\alpha_{\rm CE}$ and $\alpha_{\rm th}$ have hardly any effect on
${\rm log}F_{\rm BSS}$, since BSs from binary coalescence (not
including AML) are always much fewer than those from mass transfer
during the whole age (see the bottom panel of this figure).

The contribution from AML when ${\rm log} (t/{\rm yr})>9$ is also
presented in the bottom panel of Fig. \ref{FBSS}. We see that AML
becomes more and more important for BS formation with time, since
more and more low-mass binaries are close enough to start RLOF
with time due to AML. The contribution from AML exceeds that from
primordial binary evolution when ${\rm log} (t/{\rm yr})>9.4$.
Thus, AML in low-mass binaries should not be ignored when we study
BSs from binaries in old clusters.

\subsection {Contribution to ISED}
\begin{figure*}
\centerline{\psfig{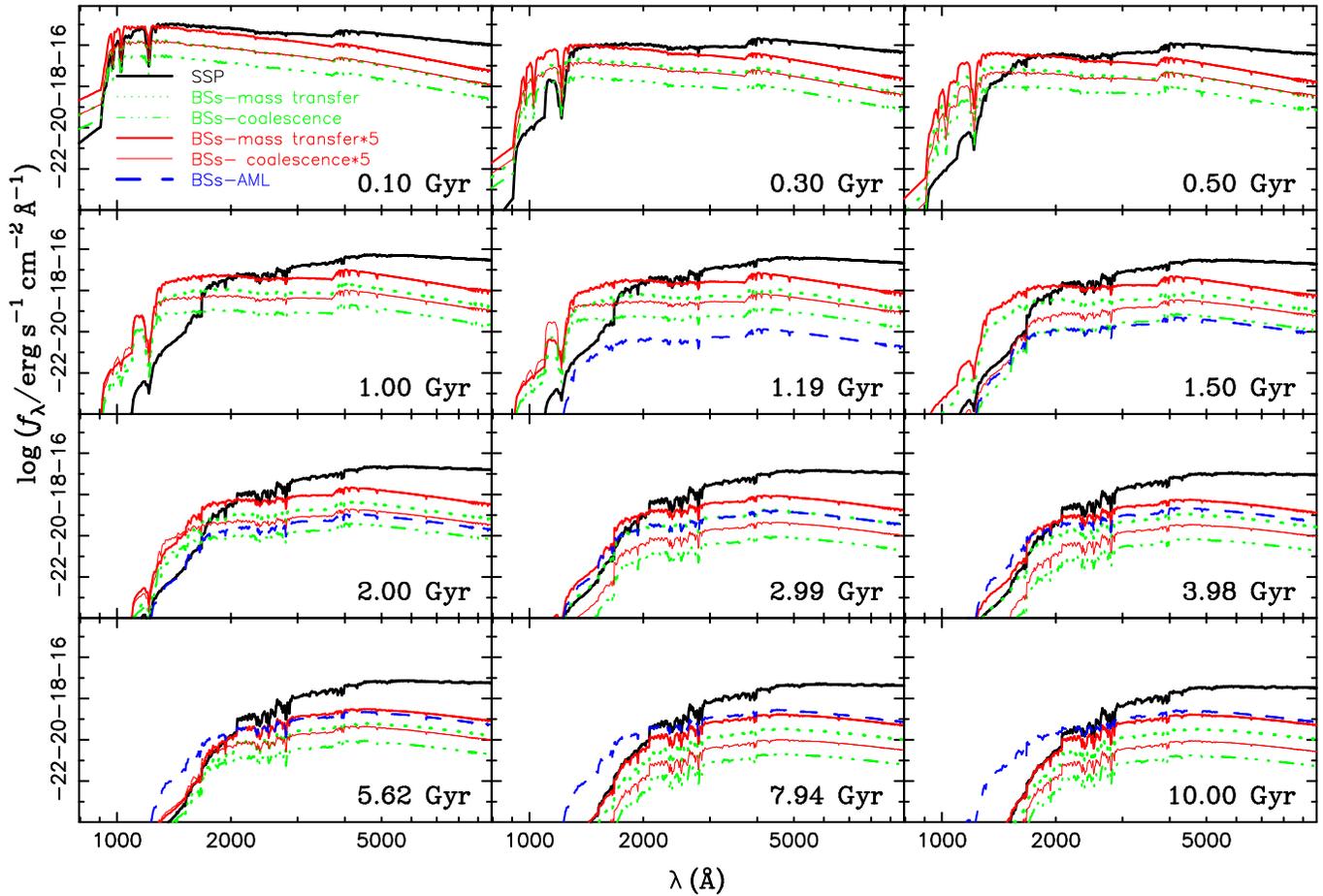}}
\caption{Integrated rest-frame intrinsic SEDs for a stellar
population (including binaries) with a mass of $1M_\odot$ at a
distance of 10kpc for set 1 with $\beta =1$ when the primary is in
HG at the onset of RLOF. The black solid lines are for the results
of SSP, which means a population without binary interaction, and
the others are the contributions of BSs from different
evolutionary channels. Note that the red solid lines are 5 times
contributions of BSs from mass transfer (the thick ones) and from
binary coalescence (the thin ones). We see that BSs resulting from
PBE (without AML) are dominant contributors to the ISED in UV and
blue bands between 0.3 and 2.0 Gyr. The BSs and SSP have
comparable energy in UV and blue bands between 2 and 4 Gyr. The
contribution from AML becomes more important with time and exceeds
that from primordial binary evolution when the population is older
than 3 Gyr.} \label{sed}
\end{figure*}

To demonstrate the contribution of BSs from PBE to ISEDs, we
plotted ISEDs of a population for various cases in Fig. \ref{sed},
where SSP means a population without binary interaction. This
figure shows that BSs resulting from binary evolution are dominant
contributors to the ISED in UV and blue bands between 0.3 and 2.0
Gyr. The BSs and SSP have comparable energy in UV and blue bands
between 2 and 4 Gyr. The contribution from AML becomes more
important with time, and exceeds that from primordial binary
evolution for a population older than $\sim 3$ Gyr. Thus,
primordial binaries are important contributors to BSs over the
whole age range.

In most cases, binary coalescence is much less important than mass
transfer, as shown in Fig.\ref{sed}. However, binary coalescence
may produce BSs with high masses, which then have lower magnitudes
i.e they may be about 2.5 magnitude brighter than the turnoff (see
Figs.\ref{m67} and \ref{nd}). Once these massive BSs are produced,
they become the dominant contributors to the ISEDs in the UV, even
if there is only one.

\subsection {Influences of the mass transfer efficiency}
The mass transfer efficiency, $\beta$, is defined as the mass
fraction of the matter lost from the primary accreted by the
secondary. In the above study, $\beta$ is set to be 0.5 except for
case A binary evolution. In fact, this value is unclear at
present. It might be higher when the mass donor is in HG, while
lower when the mass donor is on FGB or AGB.  What is the result if
we change this value, e.g $\beta =1$ when the primary is in HG at
the onset of RLOF?

Obviously, the secondary may accrete more material from the
primary, and then is more likely to become a BS. From this view,
it seems that we may obtain more BSs when $\beta =1$. However here
we ignored another important fact, that is, the lifetime of the
secondary on the main sequence after RLOF. We now consider a
simple case: the turnoff of a cluster $M_{\rm to}=1M_\odot$, and
$q_{\rm c}=3.2$ when the mass donor is on the MS and in HG. In
this case, the initial mass of the secondary $M_{\rm 2i}$ should
be less than 1$M_\odot$ to ensure the secondary to become a BS,
and binaries with $M_{\rm 1i}$ less than 3.2$M_\odot$ possibly
contribute to BSs via stable RLOF. We firstly see the result of
binaries with the primary having initial mass of $M_{\rm
1i}=1.6M_\odot$, the final mass of which is approximately assumed
to be $0.6M_\odot$. This assumption is appropriate according to
the initial-final mass relation of Han, Podsiadlowski \& Eggleton
(1994). The primary then lost $1M_\odot$ during RLOF. The minimum
mass of the secondary for stable RLOF is $M_{\rm 2i}^{\rm
min}=1.6/3.2=0.5M_\odot$. As a consequence, the secondary mass
would range from 1 to $1.5M_\odot$ for $\beta =0.5$, and from 1.5
to $2M_\odot$ for $\beta =1$ after RLOF. This means that all the
possible BSs from binaries with $M_{\rm 1i}=1.6M_\odot$ (via
stable RLOF) have already been counted when $\beta =0.5$. When we
adopted $\beta =1$, we only increased the masses of the products
while not the number. As well, the products from $\beta =1$ have
much less lifetimes on the main sequence in comparison to those
from $\beta =0.5$. Thus, after some time, the number of BSs from
binaries with $M_{\rm 1i}=1.6M_\odot$ when $\beta =1$ will be less
than that of $\beta =0.5$. The binaries with $M_{\rm 1i}$ between
1.6 and 3.2$M_\odot$ have similar results like this. Another case
is that, when $\beta =0.5$, some secondaries cannot increase their
masses to become BSs while can when $\beta =1$ (i.e $M_{\rm
1i}<1.6M_\odot$), leading to an increase of the BS number. In this
case, if $\beta =1$, the original BSs from $\beta =0.5$ will be
more massive and have shorter lifetimes on the main sequence. The
final number of BSs is related to the completion of the two facts.
We examined set 1 with $\beta =1$ and found that the BS number
decreased while $N_2$ increased in comparison to those of $\beta
=0.5$. The specific frequency of BSs when $\beta =1$ is then
smaller than that of $\beta =0.5$.

\begin{figure*}
\centerline{\psfig{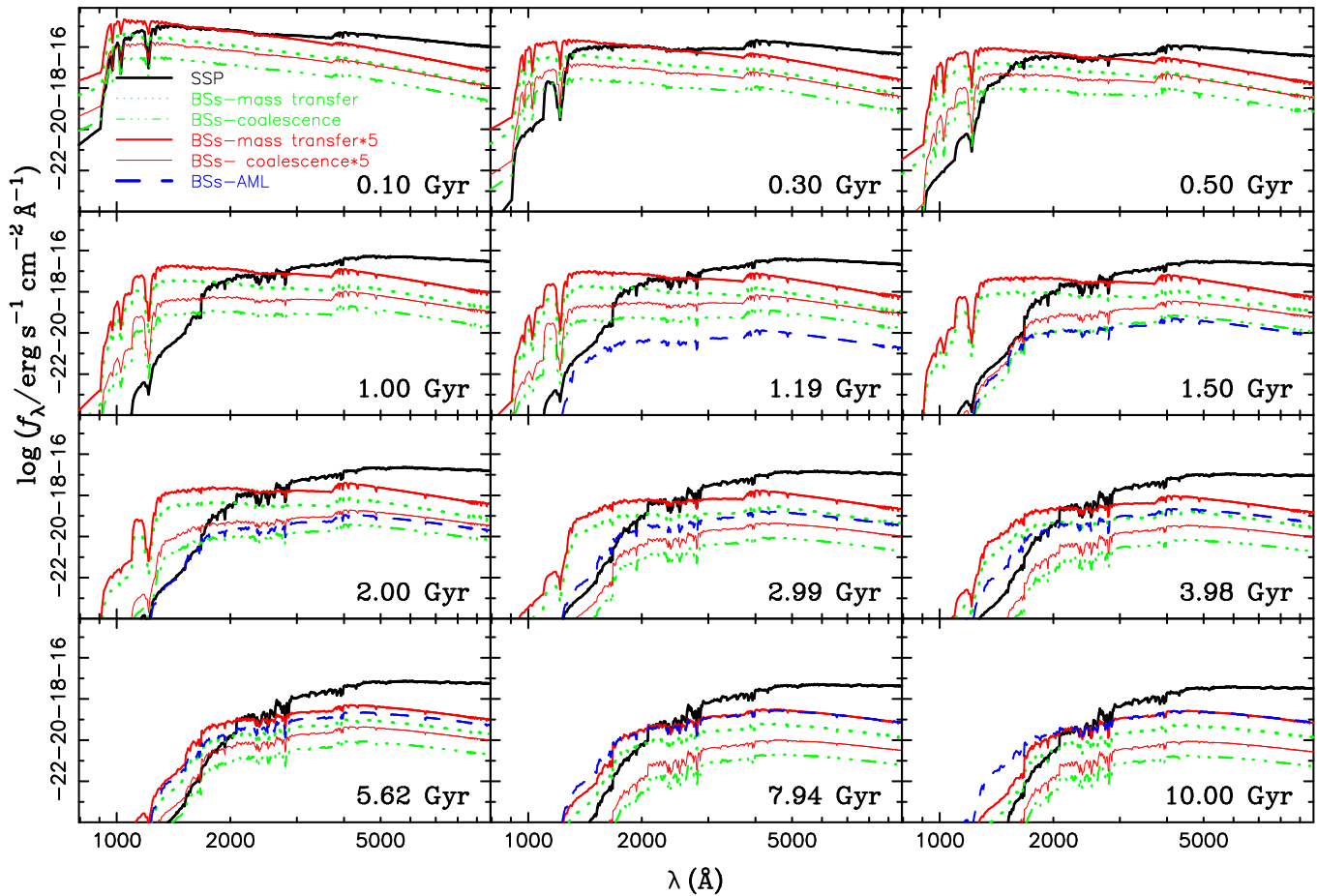}}
\caption{Integrated rest-frame intrinsic SEDs for a stellar
population (including binaries) with a mass of $1M_\odot$ at a
distance of 10kpc for set 1 with $\beta =1$ when the primary is in
HG at the onset of RLOF . The black solid lines are for the
results of SSP, which means a population without binary
interaction, and the others are the contributions of BSs from
different evolutionary channels. Note that the red solid lines are
5 times contributions of BSs from mass transfer (the thick ones)
and from binary coalescence (the thin ones). } \label{sed1}
\end{figure*}

However it should be noticed that BSs from $\beta =1$ may be more
massive than those of $\beta =0.5$, which may explain some BSs
with far away from the turnoff (see the right panel of Fig.1),
especially, we may obtain BSs with masses larger than 2 times of
the turnoff even these objects have very short lifetimes. For old
clusters, since the binaries contributing to BSs are less massive
than those in young clusters, it is then appropriate to assume a
high value of $\beta$. We will see in section 4 that the
assumption of $\beta =1$ matches the observations better than that
of $\beta =0.5$ for King 2, NGC 188, IC 1369 and NGC 2682. Since
BSs from a high $\beta$ have higher masses, their contributions to
ISED then become more important in ultraviolet and blue bands.
Fig.\ref{sed1} shows the ISEDs for the same population as that in
Fig.\ref{sed} but $\beta =1$ when the primary is in HG at the
onset of RLOF. We found that the value of $\beta$ significantly
affects the BS contribution to ISEDs, that is, a large $\beta$
results in more contribution in ultraviolet and blue bands. When
$\beta =1$, BSs from PBE dominate the ISEDs in ultraviolet and
blue bands till 5.6Gyr, after which their contributions become
smaller than those of BSs from AML.

Correspondingly, the distribution of BSs from stable mass transfer
is changed at different values of $\beta$. The results of $\beta
=1$ are plotted in Fig.\ref{nd}, frow which we see that more BSs
with lower magnitudes (or brighter than the turnoff) are obtained
at this value.

\section {The role of PBE for producing BSs in Galactic open clusters }
We select some Galactic open clusters with metalicities similar to
that of the Sun (i.e. around 0.02) to examine BSs from primordial
binary evolution. Table 2 shows the characteristics of these
clusters and their BS information from both observations and our
simulations.

\begin{table*}
 \begin{minipage}{160mm}
 \caption{Parameters of Galactic open clusters with metallicity similar to that of the Sun
 ($Z=0.02$). The bottom part shows the clusters which have no metallicity
information available, but are rich in BSs from the new catalogue
of AL07. We assumed solar metallicity ($Z=0.02$) for these
clusters when simulating. Columns (1)-(4) are cluster name, age
[${\rm log}(t/{\rm yr})$], colour excess [$E(B-V)$] and distance
modulus (the asterisks mean distance modulus corrected for
absorption). Columns (5)-(6) are the metallicity ($Z$) and [Fe/H]
values, and columns (7) and (8) are $N_{\rm 2}$(number of stars
within 2 mag below the main sequence turnoff) and $N_{\rm BS}$
(numbers of BSs) in the sample clusters from observations (the
data in brackets are from Xin et el.(2007)). Columns(9)-(11) show
our simulation results from simulation set 1: $f_0=N_{\rm bs}^{\rm
b}/N_{\rm 2}^{\rm b} \cdot N_{\rm 2}$, is the BS number expected
from PBE. $f_0^{'}$ is similar to $f_0$ but for that of AML, which
is only shown for clusters older than 1Gyr. $f=f_0/N_{\rm bs}$, is
the ratio of the BS number expected from PBE to that of observed.
As a comparison, we also show $f$ from set 2 (which obviously
differs from the other sets as presented in Fig.4) in brackets of
column(11). The last column is the references for the data (note
that the values of $N_2$ and $N_{\rm BS}$ in all clusters come
from AL07): (1)AL07; (2)Vallenari et al. 2000; (3)Manteiga et al.
1995; (4)Bragaglia et al. 2000; (5)Claria et al. 2003; (6)Hebb et
al. 2004;  (7)van Leeuwen 1999; (8)Krusberg \& Chaboyer 2006;
(9)Kharchenko et al. 2005; (10)Loktin \& Martkin 1994; (11)Dias et
al. 2002; (12)Marshall et al. 2005; (13)Lapasset et al. 2000; (14)
Vandenberg \& Stetson 2004; (15)Eggen 1981; (16)Sarajedini et al.
2004; (17)Luck 1994; (18)Paunzen \& Maitzen 2001; (19)Prosser et
al. 1996; (20)Sagar et al. 2001; (21)Surendiranath et al. 1990;
(22)Carraro \& Patat 2001; (23)Balona \& Laney 1995; (24)Ahumada
2005; (25)Andreuzzi et al. 2004.}
 \label{tab2}
   \begin{tabular}{lccccccccccc}
\hline
Cluster Name &${\rm log}(t/{\rm yr})$&$E(B-V)$&DM&$Z$& [Fe/H]&$N_{\rm 2}$&$N_{\rm BS}$&$f_0$&$f_0^{'}$& $f$& References\\
(1)&(2)&(3)&(4)&(5)&(6)&(7)&(8)&(9)&(10)&(11)\\
\hline
IC 166 &9.00 &0.80 &15.65&0.02&0.00& 120(110) & 7(11)& 1.36 &-& 0.194 (0.103)&1,2,3\\
IC 1311&8.95 &0.45 &$14.10^{*}$&0.02&0.00& 220(100) & 12(7)& 2.79&-& 0.232 (0.100)&1,4\\
IC 2488&8.25 &0.24 &11.20&0.019&-0.02& 30(10)& 3(1) &0.58 &-& 0.192 (0.0511)&1,5\\
IC 4756&8.90 &0.23 &$7.60^{*}$&0.022&0.04 & 80(55)& 6(1) &1.07 &-& 0.178 (0.0690)&1,6\\
Melotte 111&8.60 &0.00 &$4.77^{*}$&0.019&-0.03&10(10) &1(1)&0.16&-& 0.160 (0.0452)&1,7\\
NGC 188 & 9.85 &0.08 &11.35&0.019&-0.01&185(170)& 24(20)& 0.72&3.78& 0.0300 (0.0150)&1,8\\
NGC 1027&8.55  &0.33 &10.46&0.023&0.06&40(40) & 0(2)&0.66&-&-&1,9,10\\
NGC 2287&8.39 &0.03 &9.30&0.022&0.04&30(30) & 2(3)& 0.52&-& 0.258 (0.0705)&1,9,11\\
NGC 2301&8.31 &0.03 &9.76&0.023&0.06& 20(15)& 0(1)&0.35&-&-&1,9,11\\
NGC 2437&8.39 &0.15 &11.16&0.023&0.06&80(70)& 7(5)&1.38&-& 0.196 (0.0540)&1,9,11\\
NGC 2539&8.80 &0.06 &10.60&0.018&-0.04& 100(20)&1(1) &1.37&-& 1.371 (0.55)&1,12,13\\
NGC 2660&9.00 &0.40 &$12.20^{*}$&0.02&0.00&150(110)& 8(18)&1.70&-& 0.213 (0.113)&1,4\\
NGC 2682&9.69 &0.038&9.65&0.018&-0.04& 175(200)&30(30) &0.72&2.31& 0.0241 (0.013)&1,14\\
NGC 3532&8.54 &0.04 &8.59&0.019&-0.02& 160(90)& 5(9)&2.55&-& 0.511 (0.169)&1,15,16\\
NGC 5316&8.19 &0.27 &11.26&0.019&-0.02&10(20) & 0(4)&0.22&-&-&1,9,11\\
NGC 6067&8.11 &0.32 &$11.17^{*}$&0.021&0.01&25(60)& 3(7)&0.60 &-&0.201 (0.035)&1,17\\
NGC 6281&8.51 &0.15 &8.93&0.02&0.00&20(25)& 0(4)&0.32&-&-&1,9,11\\
NGC 6475&8.34 &0.07 &7.30&0.022&0.03&10(15)&2(2)&0.16&-& 0.08 (0.0219)&1,11,19\\
NGC 6940&8.94 &0.21 &$10.08^{*}$&0.021&0.01&90(130)&6(7)&1.16&-&0.193 (0.0796)&1,9,11\\
&&&&&&&\\
IC 1369&9.16 &0.57 &$11.59^{*}$&0.02?&...&35(35) &6(6) &0.35&0.03&0.058 (0.0226)&1,10,11\\
King 2 &9.78 &0.31 &$13.80^{*}$&0.02?&...&250(250)&30(30) &0.97&4.01&0.032 (0.0159)&1,11\\
Melotte 105&8.40 &0.52 &$11.80^{*}$&0.02?&...&80(25)& 4(1)&1.37&-& 0.341 (0.0923) &1,20\\
NGC 2818&8.70 &0.22 &$12.90^{*}$&0.02?&...&45(45)& 2(2)&0.70&-& 0.347 (0.104)&1,21\\
NGC 3114&8.48 &0.07 &$9.80^{*}$&0.02?&...&14(50) & 5(5)&0.23&-& 0.046 (0.0138)&1,22\\
NGC 3496&8.78 &0.52 &$10.70^{*}$&0.02?&...&35(70)& 1(4)&0.52&-&0.515 (0.129)&1,23\\
NGC 6416&8.78 &0.25 &$10.12^{*}$&0.02?&...&50(35)& 1(3)&0.74&-&0.735 (0.185) &1,9\\
NGC 6939&9.11 &0.34 &$11.30^{*}$&0.02?&...&180(80) & 5(4)&1.77&0.06& 0.354 (0.161)&1,25\\
\hline \label{clusters}
\end{tabular}
\end{minipage}
\end{table*}

In the new BS catalogue of Ahumada \& Lapasset \shortcite{al07}
(hereafter AL07), the authors showed a schematic diagram for the
regions of BSs on a colour-magnitude diagram to identify BSs. They
have not given a definitive area for BSs on a CMD. In this paper,
we simply define BSs as main sequence stars which have masses
larger than the turnoff mass of the host clusters. We will see
later that BSs defined in this way are well matched with the
region defined in AL07.

To demonstrate the importance of PBE for BS formation in a
cluster, we introduce a factor, $f$, which is defined as:
\begin{equation}
f=N_{\rm bs}^{\rm b}/N_{\rm 2}^{\rm b} \cdot N_{\rm 2}/N_{\rm bs},
\end{equation}
where $N_{\rm bs}^{\rm b}$ and $N_{\rm 2}^{\rm b}$ are the numbers
of BSs and stars within a 2 mag interval below the main-sequence
turnoff point of the host cluster from the simulations,
respectively, and $N_{\rm bs}$ and $N_{\rm 2}$ are from
observations. Thus, $f$ is actually the ratio of the BS number
expected from PBE to that observed. The values of $f$ from
different calculations are also presented in Table 2.

\begin{figure*}
\centerline{\psfig{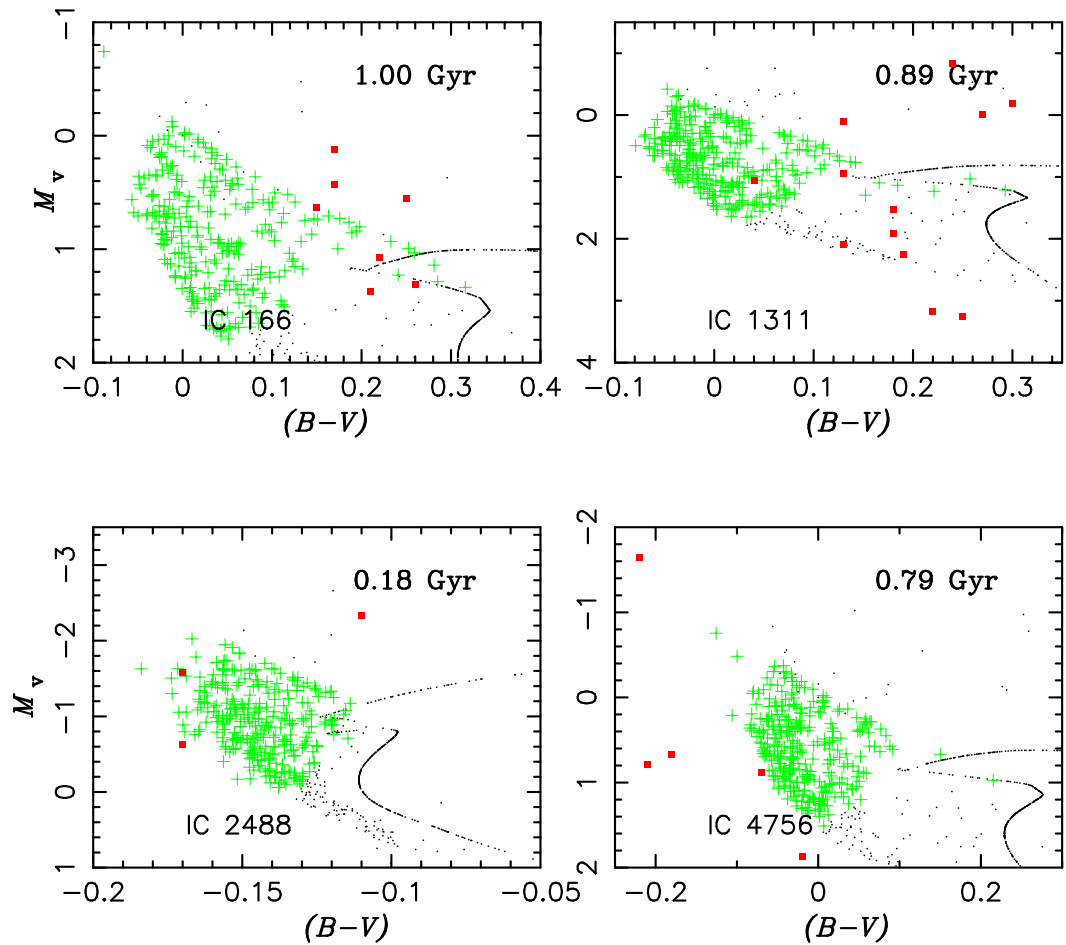}}
\caption{Colour-magnitude diagram for IC 166, IC 1311, IC 2488 and
IC 4756. The crosses and squares are BSs from PBE and AL07,
respectively. The dots are for all other stars in the cluster from
our simulation.} \label{cmd1}
\end{figure*}

\begin{figure*}
\centerline{\psfig{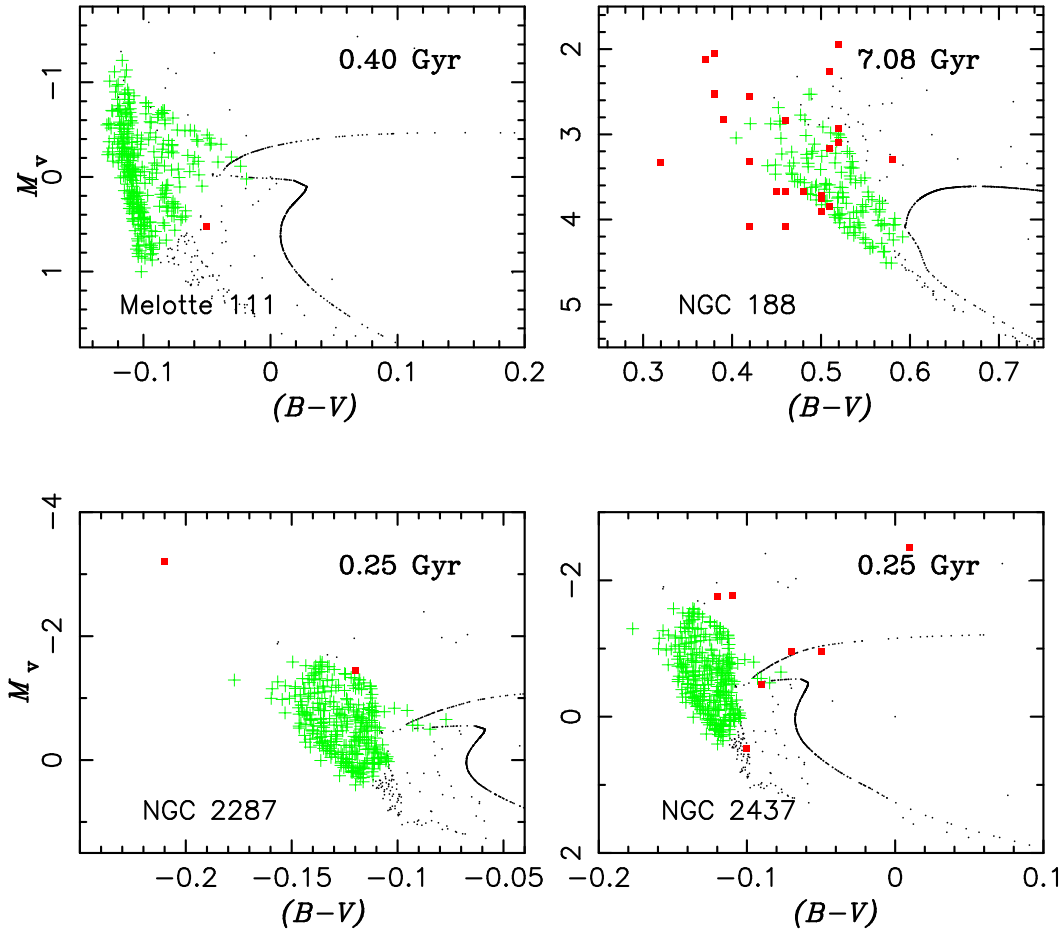}}
\caption{Similar to \ref{cmd1}, but for Melotte 111, NGC 188, NGC
2287 and NGC 2437.} \label{cmd2}
\end{figure*}

\begin{figure*}
\centerline{\psfig{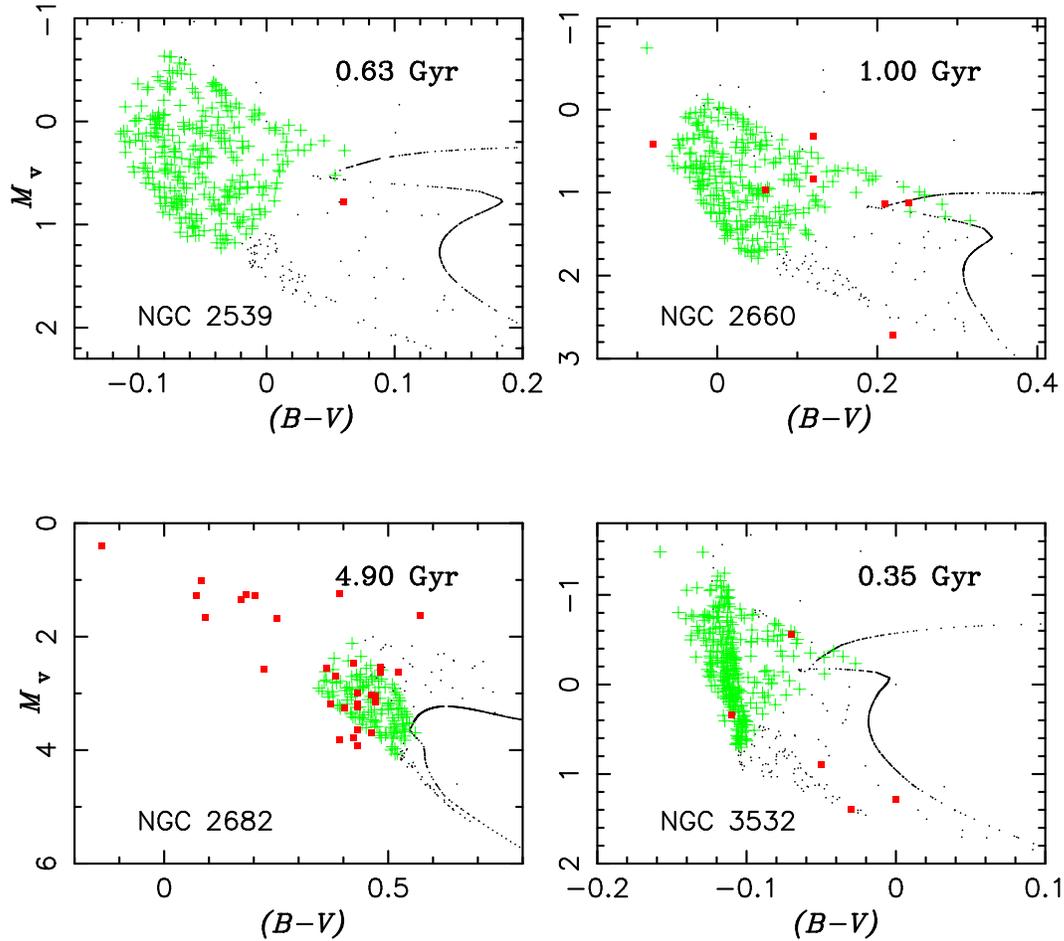}}
\caption{Similar to \ref{cmd1}, but for NGC 2539, NGC 2660, NGC
2682 and NGC 3532.} \label{cmd3}
\end{figure*}

\begin{figure*}
\centerline{\psfig{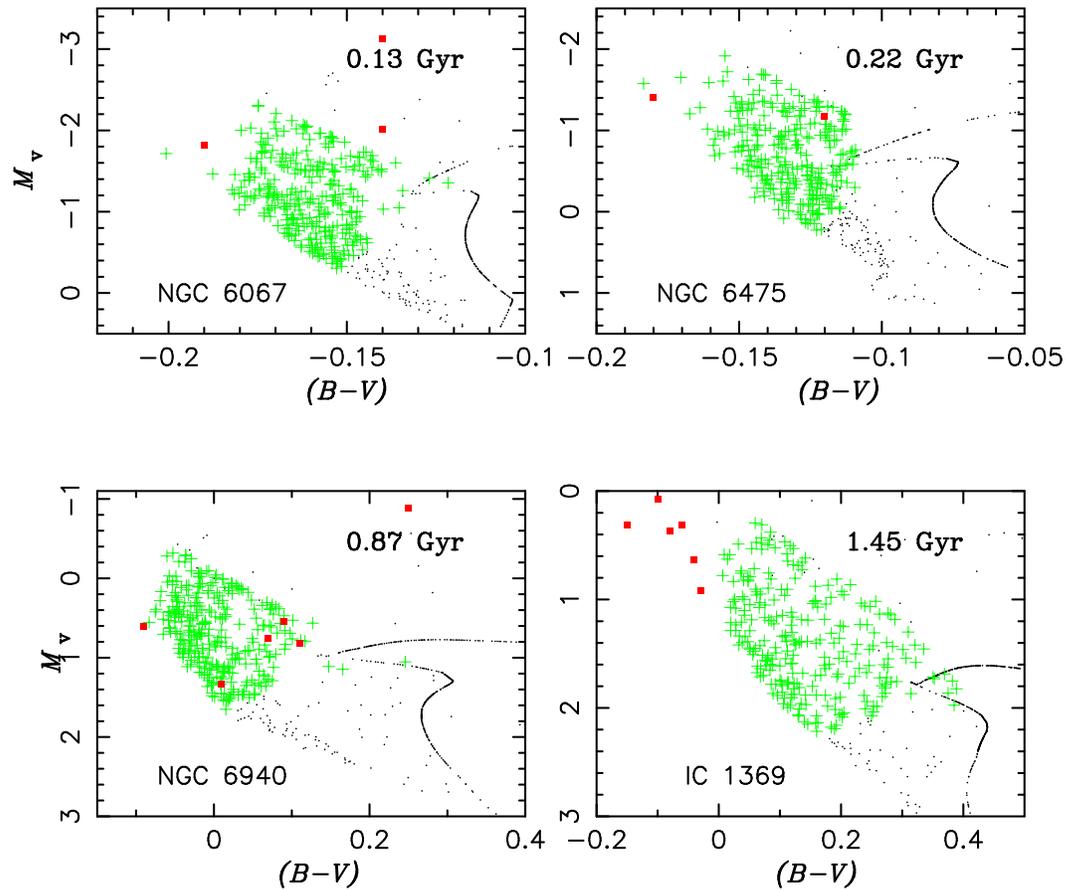}}
\caption{Similar to \ref{cmd1}, but for NGC 6067, NGC 6475, NGC
6940 and IC 1369.} \label{cmd4}
\end{figure*}

\begin{figure*}
\centerline{\psfig{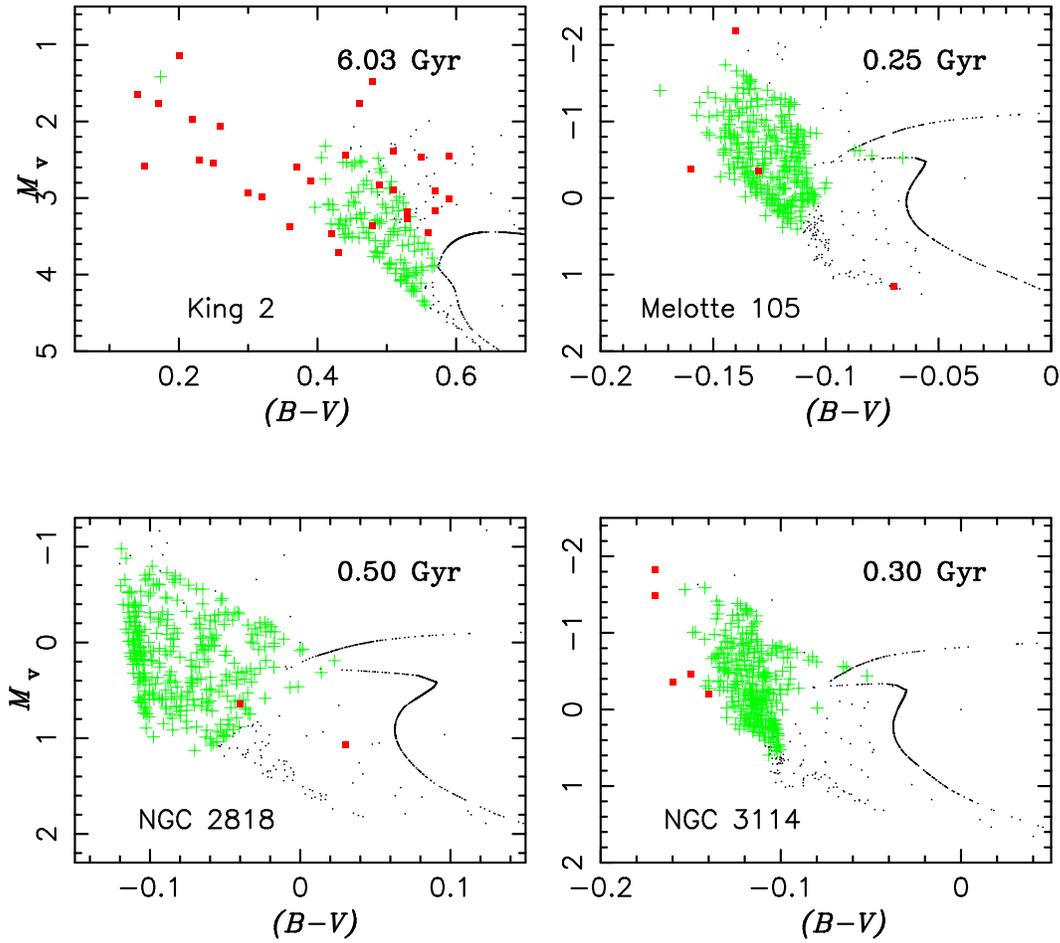}}
\caption{Similar to \ref{cmd1}, but for King 2, Melotte 105, NGC
2818 and NGC 3114.} \label{cmd5}
\end{figure*}

\begin{figure*}
\centerline{\psfig{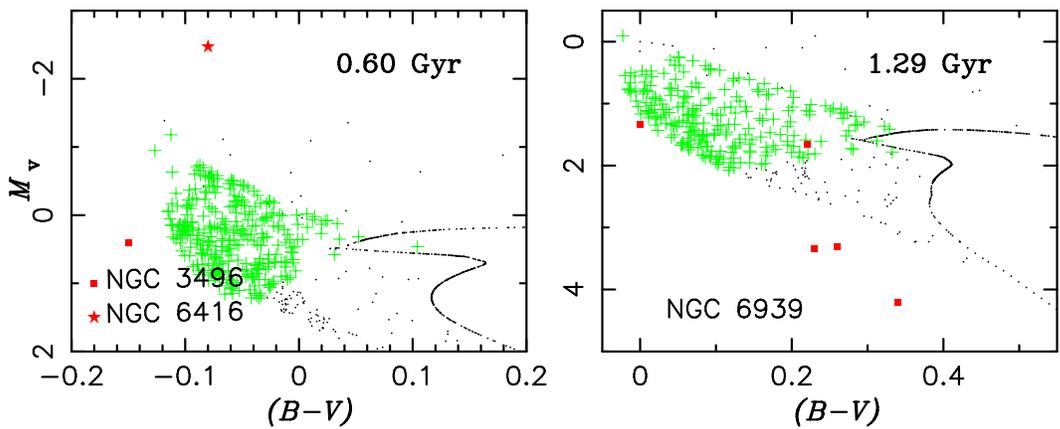}}
\caption{Similar to \ref{cmd1}, but for NGC 3496, NGC 6416 and NGC
6939. Note that the results of NGC 3496 and NGC 6416 are in the
same panel (in the left) since they have the same age, which
results in the same CMD from our simulation.} \label{cmd6}
\end{figure*}

Figs. \ref{cmd1} to \ref{cmd6} present colour-magnitude diagrams
for the sample clusters in Table 2 (the clusters appear in the
same order as that in Table 2 in these figures), where the crosses
are BSs from PBE and the squares are observed BSs. We have not
shown the clusters for which no BSs were presented in AL07.

\subsection{Discussions on the sample clusters}
From Figs.\ref{cmd1} to \ref{cmd6} we see that some observed BSs
are in the appropriate regions obtained from PBE, while some are
not. The outlying BSs may be divided into four cases as
follows\footnote{Note that even though the BSs locate in the
appropriate regions, it is also possible to produce them in other
ways. For instance, BSs from AML are also around the turnoff of
the cluster \cite{chen08a}.}.

\begin{figure*}
\centerline{\psfig{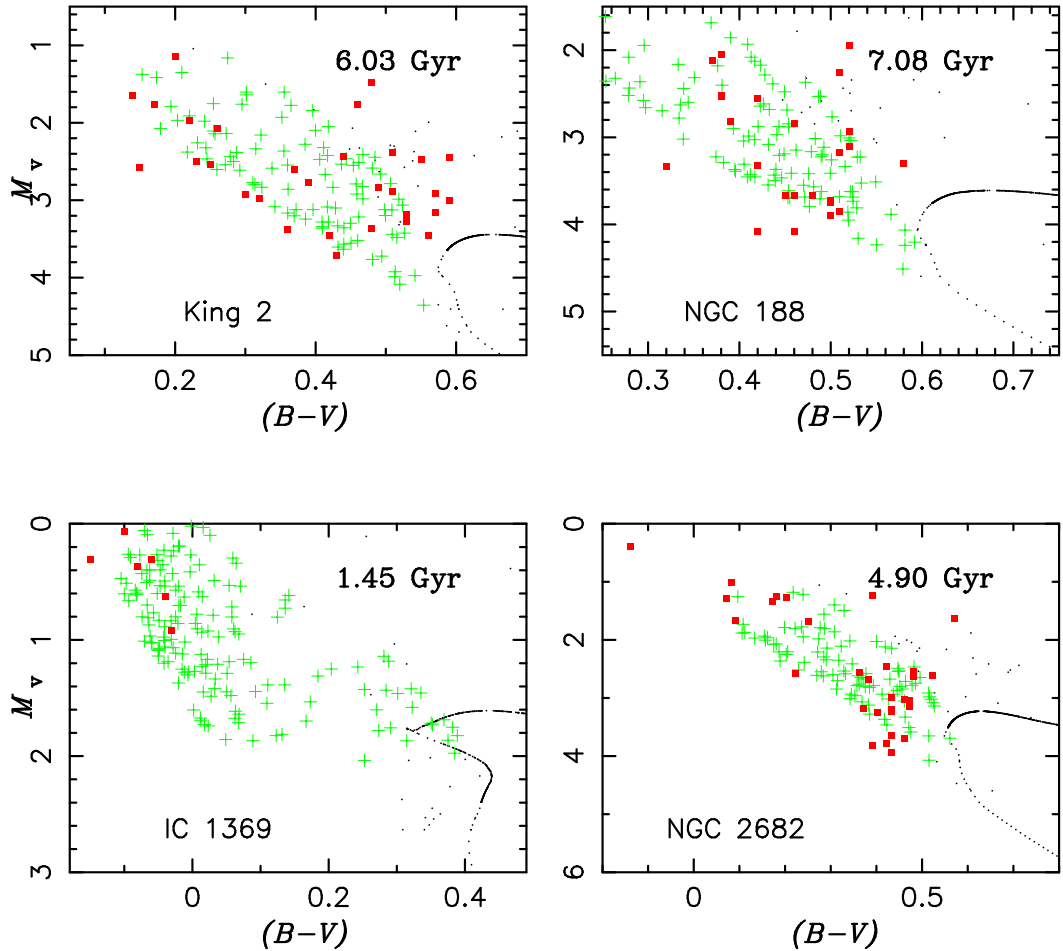}}
\caption{Similar to \ref{cmd1}, but for King 2, NGC 188, IC 1369
and NGC 2682 and the mass transfer efficiency $\beta =1$.}
\label{cmd7}
\end{figure*}

(i) Some observed BSs are much more luminous in the V band than
the turnoffs of the host clusters, such as, in IC 1369, IC 4756,
King 2, NGC 188, NGC 2287, NGC 2682, NGC 6067 and NGC 6416.  These
luminous BSs are likely from {\it Channel II} except for the most
luminous BS in NGC 2682 (F81), but this channel seems to have an
effect only in King 2 from the CMDs of these clusters. We know
that {\it Channel II} is very sensitive to the cluster age (see
Section 3.1), but the age determination of a cluster is very
uncertain at present. For instance, the age of NGC 2682 may range
from 3 \cite{bb03} to 6 Gyr \cite{jp94}. Thus, it is still
possible that {\it Channel II} leads to the production of these
luminous BSs. Another possibility is that the mass fraction of the
matter lost from the primary and accreted by the secondary
($\beta$) is more than 50 per cent when the primary is in HG.
Obviously, we may obtain more luminous objects with a high
$\beta$. As mentioned in section 3.4, it is appropriate to assume
a high value of $\beta$ in old clusters. We re-plotted CMDs with
$\beta =1$ for clusters King 2, NGC 188, IC 1369 and NGC 2682 in
Fig.\ref{cmd7}, from which we see that the assumption of $\beta
=1$ matches the observations better than that of $\beta =0.5$ for
the four clusters.

(ii) Some BSs are redder than those obtained from PBE, such as in
IC 166, IC 1311, IC 2488, Melotte 105, NGC 2437 and NGC 6940 (also
in King 2, NGC 188 and NGC 2682). These objects are likely
sub-giant stars from their locations on the CMDs. Thus, they are
ruled out by the BS definition (only stars on the main sequence
are included) in this paper.

(iii)Several BSs from AL07 are below the turnoffs, but bluer than
normal stars on the main sequence with the same masses\footnote{
Here `normal' means that the stars have not passed through RLOF,
during which a low-mass main sequence star may increase its mass
by accretion. After RLOF, the accretor is rejuvenated and bluer
than a normal star with the same mass.}, such as in IC 1311,
Melotte 105, Melotte 111, NGC 2539, NGC 2660, NGC 2818, NGC 3532
and NGC 6939. These objects are inside the region defined in AL07,
but have masses lower than the turnoffs, thus, they are also ruled
out by the BS definition in this paper.

(iv) Several observed BSs are bluer than those from PBE, such as
in IC 4756, NGC 188, NGC 3114 and NGC 3496. Enhancement of surface
helium abundance of BSs from {\it Channels I and II} may lead them
to be bluer than those shown in the figures, but the blueshift is
within 0.1 mag, as shown in Chen \& Han
\shortcite{chen04,chen08a}. BSs which are bluer beyond 0.1 mag
than the left side of the region from PBE (i.e in IC 4756) could
have larger He abundances than the examples shown in Chen \& Han
\shortcite{chen04,chen08a}. AGB stars can provide material with
high He abundances, but from the simulations this only occurs in
populations older than 2.24 Gyr for Population I. Thus, it is
possible that these BSs originate from mass transfer between AGB
and MS stars, and have long-orbital periods in old clusters.
However, it is still a puzzle in young open clusters -- where does
the extra He come from? It is possible that the merger models in
Chen \& Han \shortcite{chen08a} should be revised, e.g. the mass
fraction of the primary mixed with the secondary is less than that
assumed in that paper, but this fraction itself is now an open
problem.

Many BSs observed in  NGC 188, NGC 2682 and King 2 are in the
regions from PBE, but we can obtain only one in each of the three
clusters if we normalize BS numbers to $N_{\rm 2}$ (see Table 2).
All of the three clusters are old open clusters and AML of
low-mass binaries are much more important for BS formation than
PBE in them. However, as shown in Table 2, although the number of
BSs from AML may be as much as 4, AML also seems to be unable to
explain the observed BS numbers in these three clusters.
Meanwhile, AML should have no influence on BS formation in young
open clusters, but we see from Table 2 that $f$ is generally
around 0.2 (much less than unity) in clusters younger than 1 Gyr,
except for NGC 3114 ($f=0.046$) and NGC 6475 ($f=0.08$). The low
BS specific frequency (less than 50 per cent in most cases) from
our simulation for most clusters seems to support other channels
producing BSs in addition to PBE even in young open clusters. It
is puzzling since PBE is expected to dominate BS formation in open
clusters younger than 1 Gyr (see discussions in section 4.2).

\subsection{Possible causes for the low specific frequency in the sample clusters }
It is a serious problem that the BS specific frequency obtained in
our simulations is much lower than that observed. Our first
thought is to check the validity of the result. We examined the
cases for ${\rm log}t=9$ from Hurley's rapid binary evolution code
\cite{hur02} based on the assumptions of simulation set 1 in this
paper, and did find a higher value of $f$, i.e. $f=0.3468$ for IC
166 and $f=0.3796$ for NGC 2660. We checked and found that this is
due to the different calculations of $N_2$ in the two codes. In
Hurley's code, a binary is treated as a single star, the SED of
which comes from the contribution of two components, while in
Han's code, a binary is assumed to be two single stars, since the
evolution of the primary after RLOF has not been followed.
However, the values of $f$ in Table 2 are still much less than
that expected if we multiply by a factor of 2. Following are some
discussions on this problem.

\begin{figure*}
\centerline{\psfig{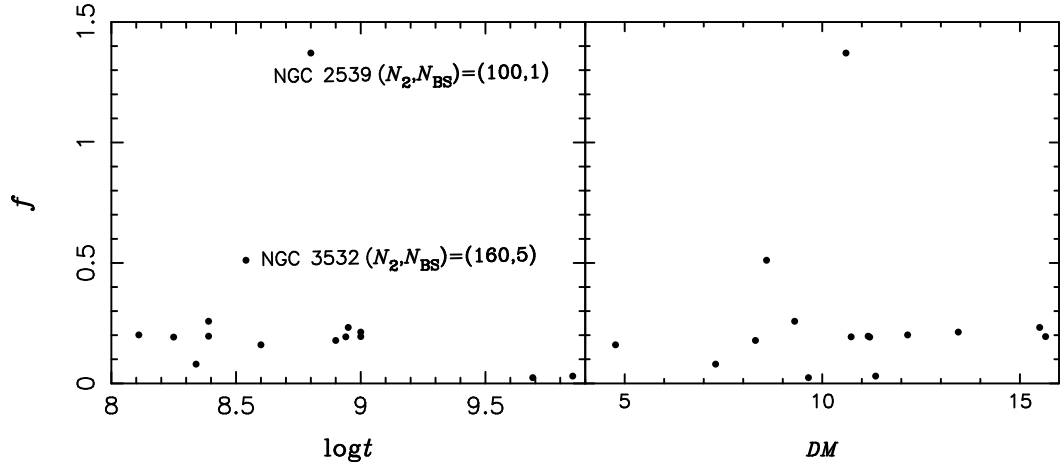}}
\caption{The ratio of the BS number expected from PBE to that of
observed, $f$, verus the cluster age and distance modulus in the
sample clusters with known metallicity. Two clusters, NGC 2539 and
NGC 3532, have $f$ obviously larger than that in the others. We
show the values of $N_2$ and $N_{\rm BS}$ for the two clusters
from AL07. } \label{f}
\end{figure*}

The value of $f$ strongly depends on $N_2$ counted in clusters --
a large $N_2$ leads to a higher $f$ from eq(9), but for some
clusters the observed values of $N_2$ are significantly different
from various authors . For comparison, we show the values of $N_2$
and $N_{\rm BS}$ from Xin et al. (2007)(Xin07) in Table 2. We see
that, for NGC 2539, the value of $N_2$ of AL07 is 5 times that of
Xin07, leading to a factor of 5 difference of $f$. The value of
$N_2$ from observation may be relevant to the cluster age and the
distance modulus. For instance, in a distant cluster, stars are
more difficult to detect, resulting in a smaller $N_2$, and thus a
smaller $f$. Also with increasing age, stars within 2 mag below
the main-sequence turnoff become less massive and more difficult
to detect, leading to a smaller $f$ as well. Fig. \ref{f} shows
$f$ varying with age and distance modulus for the sample clusters
with known metallicity. However, we have not found correlations
between $f$ and the cluster age when ${\rm log}t\le 9$ or between
$f$ and the distance modulus. When ${\rm log}t > 9$, the value of
$f$ sharply decreases due to the decrease of $N_{\rm BS}^{\rm b}$
and the increase of $N_{2}^{\rm b}$(see section 3.2). Two
clusters, NGC 2539 and NGC 3532, have $f$ obviously larger than
that for the others, due to the large $N_2$ from AL07. If we adopt
the results of Xin07, $f$ in these two clusters is much smaller
and is similar to that in the other clusters.

From section 3.2 we know that the specific frequency of BSs
depends heavily on the age. Thus, the precision of age
determination also has effect on $f$, but the cluster age is very
uncertain from observations. For instance, the age of NGC 2682
ranges from 3 \cite{bb03} to 6Gyr \cite{jp94}, which leads $f$ to
change from 0.0312 to 0.0231, about a factor of 1.35 difference.

Another possibility for the low $f$ in our simulations is that the
adopted Monte Carlo simulation parameters (e.g initial mass
function) are different from those in the clusters. The Monte
Carlo simulation parameters in this paper are for field stars and
it is very likely that these parameters are different in clusters.
For instance, the initial mass function is still an open problem
and recent studies show that it might be quite different between
disk stars and halo stars in the Galaxy \cite{pol08}. In addition,
we know that low-mass stars are located in the extended region of
the halo in old open clusters due to equipartition of energy.
These objects are then likely to increase their velocities to be
beyond the escape velocities of the host clusters and become field
stars. Thus, the number of low-mass stars (less than 1$M_\odot$)
in old open clusters is much less than that expected from the IMF
of the field stars. Moreover, both the binary frequency and the
distribution of initial orbital periods of a population are
important for BS production, and they are related to star
formation history and environment of a cluster. Thus, they might
be different even among open clusters.

Finally, the low $f$ in our simulations might indicate other
channels (e.g a more recent era of star formation and dynamical
interaction, see a review of Stryker 1993) to produce BSs in open
clusters in addition to PBE and AML. Figure \ref{f} shows that $f$
in some clusters is lower than that in the others. Thus, we cannot
rule out other channels in these clusters, although we can
multiply by a factor to increase $f$ due to the causes above.

{\it Discussions on binaries with a hot WD and a low-mass
main-sequence star}. What do these systems look like? Since the
luminosity of a hot WD is much higher than that of main-sequence
stars, it seems very likely that these systems are in the BS
region. If so, we lose BSs like this in our simulations as we have
not considered the contribution of the primaries after RLOF
(initially hot WDs) to the SED. However, we should bear in mind
that a hot WD mainly radiates in UV or far-UV, completely
different from that of a low-mass main-sequence star. For example,
a $0.57M_\odot$ CO WD is about 130000 K at 0.1Myr after its
formation according to eq(90) in the paper of Hurley, Pols \&
Tout(2002). If we assume the WD to be a black body, the peak
wavelength of its radiation is about 227A, and the flux at the
peak wavelength is $f_{\lambda =227{\rm A}} =1.5 \times 10^{30}
{\rm erg s^{-1}}$, while $f_{\lambda}$ is only about a few
$10^{26} {\rm erg.s^{-1}}$ at U,B and V bands. As the WD cools,
the peak wavelength increases, but $f_{\rm \lambda}^{\rm max}$
decreases. So the contributions of WDs in U, B and V bands are
always much less than those of low-mass main-sequence stars,
indicating that it is impossible that binaries with a hot WD and a
low-mass main-sequence star look like BSs in the diagram of V
versus B-V of a cluster. As a test, we examined the case of
$t=9.44$ Gyr, where both of two components of a binary are
included (the primaries are WDs after RLOF), and found that all
BSs obtained in this calculation are characterized by the
main-sequence components, not the WDs.

\section{Summary}
We systematically studied the BSs originating from primordial
binary evolution (PBE) via a binary population synthesis approach,
and examined their contribution to the integrated spectral energy
of the host clusters. The study shows the following results: (a)
PBE may produce BSs at any given age and BSs from it are located
within 2.5 mag of the turnoff of the host cluster. (b) In UV and
blue bands, BSs from PBE contribute the most energy for a
population between 0.3 and 2.0 Gyr, and have a contribution
similar to that of SSP between 2 and 4 Gyr. AML becomes more and
more important for BS production with time, and its importance
exceeds that from PBE in a population older than 3 Gyr. (c) BSs
from PBE have an obvious scatter on the main sequence with
decreasing mass (BS from AML also show this feature as shown in
Fig.11 of Chen \& Han 2008a) for populations between 0.4 and 1.5
Gyr. The scatter of BSs near the turnoff has been found in several
open clusters. (d) Binary coalescence from contact binaries ({\it
Channel II} in section 2.1) is very sensitive to cluster ages, but
BSs from it have relatively high mass. Thus, once {\it Channel II}
works in a cluster, BSs from it become more important than those
from other channels to contribute to the ISED of the cluster. (e)
The value of $\beta$ significantly affects on the final results,
e.g the specific frequency of blue stragglers decreasing with
$\beta$, blue stragglers produced from a high value of $\beta$
having higher masses, then contributing more to the ISEDs of the
host clusters. For old open clusters, the assumption of $\beta =1$
when the primary is in HG at the onset of mass transfer matches
the observations better than that of $\beta =0.5$ from the
locations of BSs on the CMDs.

We know that BPS results are sensitive to the uncertainties of the
initial parameters, especially the initial mass-ratio
distribution. Meanwhile, dynamical instability criteria during
RLOF will also affect the final outcomes. Different simulation
sets in this paper show that these parameters have little
influence on the features above except for the BS specific
frequency.

The main influence on the BS specific frequency comes from the
distribution of initial mass ratio. The difference between set 1
(constant) and set 2 (uncorrected) may be up to about an order of
magnitude for populations younger than 1 Gyr. With increasing
time, the effect from dynamical instability criteria becomes
obvious (up to about 0.5 dex), since more and more binaries
undergo mass transfer between giant stars and main-sequence ones.
However, neither of the five sets in our simulations can produce
enough BSs from PBE in comparison to observations in most Galactic
open clusters, even in those younger than 1 Gyr. The ratio of the
BS number expected from PBE to that of observed is only around 0.2
for most clusters we simulated. The role of AML in old open
clusters is also less important than that of expected. It is
puzzling and has been discussed in section 4.2.

\section{ACKNOWLEDGMENTS}
We are grateful to an anonymous referee for his/her valuable
comments on our work and to Dr. Pokorny R. S. for his helpfulness
improving the language of this paper. This work was in part
supported by the Chinese National Science Foundation (Grant Nos.
10603013 and 10433030,10521001 and 2007CB815406), the Chinese
Academy of Sciences (Grant No. 06YQ011001) and Yunnan National
Science Foundation (Grant No. 08YJ041001).

\end{document}